# Transit Timing Observations of the Extrasolar Hot-Neptune Planet GL 436b

Guy S. Stringfellow*, Jeffrey L. Coughlin†, Mercedes López-Morales**, Andrew C. Becker‡, Tom Krajci§, Fabio Mezzalira* and Eric Agol‡

*Center for Astrophysics and Space Astronomy, Department of Astrophysical and Planetary Sciences, University of Colorado, Boulder, Colorado
†Department of Astronomy, New Mexico State University, Las Cruces, New Mexico
**Department of Terrestrial Magnetism, Carnegie Institution of Washington, DC
‡Department of Astronomy, University of Washington, Seattle, Washington
§Astrokolkhoz Observatory, Cloudcroft, New Mexico

**Abstract.** Gliese 436 is an M dwarf with a mass of 0.45 $M_\odot$ and hosts the extrasolar planet GL 436b [3, 6, 7, 2], which is currently the least massive transiting planet with a mass of ∼23.17 $M_\oplus$ [10], and the only planet known to transit an M dwarf. GL 436b represents the first transiting detection of the class of extrasolar planets known as "Hot Neptunes" that have masses within a few times that of Neptune's mass (∼17 $M_\oplus$) and orbital semimajor axis <0.1 AU about the host star. Unlike most other known transiting extrasolar planets, GL 436b has a high eccentricity (e∼0.16). This brings to light a new parameter space for habitability zones of extrasolar planets with host star masses much smaller than typical stars of roughly a solar mass. This unique system is an ideal candidate for orbital perturbation and transit-time variation (TTV) studies to detect smaller, possibly Earth-mass planets in the system. In April 2008 we began a long-term intensive campaign to obtain complete high-precision light curves using the Apache Point Observatory's 3.5-meter telescope, NMSU's 1-meter telescope (located at APO), and Sommers Bausch Observatory's 24" telescope. These light curves are being analyzed together, along with amateur and other professional astronomer observations. Results of our analysis are discussed. Continued measurements over the next few years are needed to determine if additional planets reside in the system, and to study the impact of other manifestations on the light curves, such as star spots and active regions.

## INTRODUCTION

Low-mass stars dominate the initial mass function, constituting the largest stellar population residing in the Galaxy. M-dwarfs far outnumber solar type stars. Detection of planets about such stars have fundamental implications on the preponderance of planetary systems in the Galaxy, and hence on the proliferation of life in the Galaxy. The vast majority of planets discovered to date orbit solar type stars. Most extrasolar planets discovered to date have masses similar to Jupiter ($\geq 0.5\ M_J$), with only a handful having significantly smaller masses that range from that of Neptune down to several times that of Earth. Interestingly, these super-Earths are predominantly orbiting low-mass M-dwarf stars. The smaller radii of low-mass stars compared to solar mass stars, along with the contrast ratio of their brightness, enable the detection of smaller (radii) planets given the photometric precision possible from ground-based telescopes. This is a significant realization because transiting light curves provide direct measurement of their mass, ra-dius, and by inference, their composition. Hot-Neptune mass planets with larger radii could retain H/He atmospheres, while those more compact could be comprised of water ices or rocky material. Thus, photometric transiting light curves are able to measure the composition of hot-Neptunes and super-Earths. We have begun a program of intensive transit monitoring of M-dwarfs to search for new transiting planets, and to search for variations in the timing of transits (TTVs) in GL 436 that would signal the presence of other unseen planets orbiting the system [1, 5, 8]. Other M-dwarf transiting systems will be added to the TTV element of the program as discovered. We report initial results of the search for TTVs of the $\sim$23 $M_\oplus$ hot-Neptune planet that transits GL 436 [3].

## TRANSIT OBSERVATIONS OF GL 436

Observations of GL 436 were taken in the V-filter on the nights of 2008UT April 7, April 28, and May 6 with the 3.5-meter telescope at the Apache Point Observatory; we were weathered out the night of May 13 at APO (see [4] for details of the reduction). Resulting individual data points have typical errors of 1-2 mmag, and a typical cadence of about 17 seconds with individual exposure times ranging from 0.8s-1.5s, depending on the nightly conditions. Figure 1 shows results for the APO 3.5m runs.

Observations supporting our 3.5m APO runs were also carried out by our team using other facilities. These included the NMSU 1m telescope located adjacent to the APO 3.5m dome, the Sommers Bausch Observatory (SBO) 24" telescope located on the campus of CU-Boulder, and the Astrokolkhoz Observatory 11" telescope located at Cloudcroft, NM (TK 11"). Results for two of the nights (April 28 and May 6) are shown in Figure 2, along with the corresponding 3.5m APO data for the same night. All these data were consistently reduced and modeled in the same manner as was done for the 3.5m data (see [4]), as were amateur data obtained through the web site of Bruce Gary (an example of which is included in Figure 2a), and previously published data.

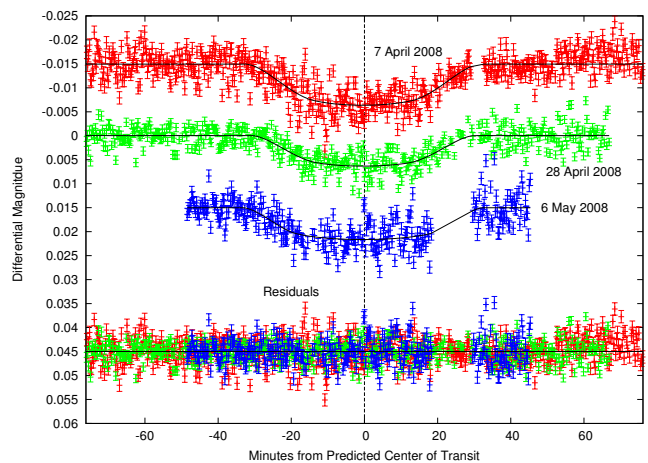

**FIGURE 1.** APO 3.5m observations of GL 436 conducted in the V-band. Transits were observed on three separate nights. Shown are unbinned data with measured error bars, computed model fits to these data, and the associated residuals.

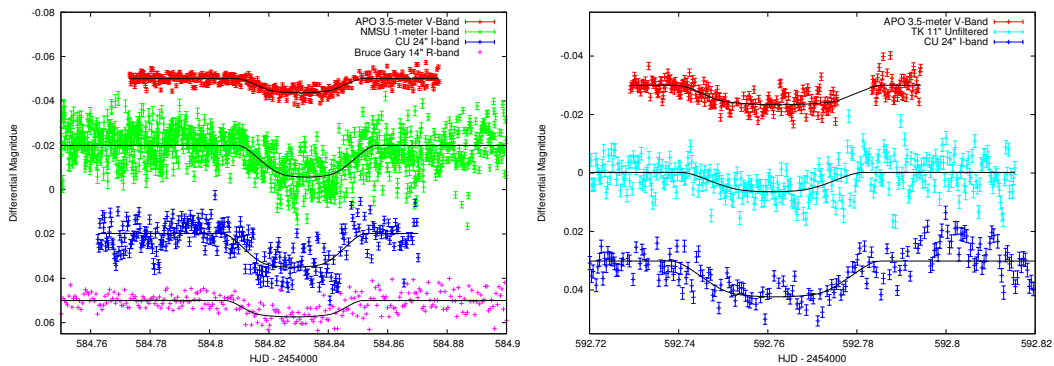

**FIGURE 2.** Collection of transiting data for GL 436 obtained on the nights of 2008 April 28 (left panel) and 2008 May 6 (right panel). The plots are color coded based on the telescope used in observing the transit. Filters used are identified in the panels. Model fits are superimposed on the unbinned data.

## MODELING AND ANALYSIS

We use the JKTEBOP code [9] to model the transit light curves (see [4] for details). Fits to the transit light curves are shown in Figures 1 and 2. Figure 3 shows the O-C diagram for transit timing variations (left panel) computed from a new ephemeris we have derived [4]. The baseline is only about a year, and there is no indication of any TTV present above a minute. The data is not yet sensitive to variations occurring on timescales less than a minute. Note the high precision of the 3.5m APO data, and the large scatter present in the amateur data − probing timing variations at this level is challenging even for the best amateur systems, and great care in the observing technique must be exercised. This became evident from our many attempts at measuring good transits with the Astrokolkhoz data, for which we were only able to obtain useable transit data after honing our technique over many previous transits (see Figure 2b).

## CONCLUSIONS

Analysis of ground-based observations presented herein indicate some fundamental properties of transiting planets can be securely deduced (such as the planet's mass, radius, and orbital inclination), but results regarding the mid-point ($T_0$), depth ($T_d$), and width ($T_w$) appear less certain amongst the various observations, perhaps conditional upon the aperture, resulting precision, and the filter bandpass used (e.g., we find $T_d$ in the I-band is roughly a factor 2 deeper than in the V-band when the ratio of planet-to-star radii is allowed to vary during fitting, but less so when fixed). $T_d$ and $T_w$ are the most problematic quantities, being highly dependent on structure (real or noise) present in the transiting light curve. If real, structure could, for example, result from the migration of spots on the surface of the star between transits. $T_d$ and $T_w$ impact the determination of $T_0$, and (through the latter) the search for TTVs. Nevertheless, all the data presented herein detect the hot-Neptune planet orbiting GL 436 and render consistent derivations of the planets basic properties (see Figure 3, right panel). Transit observations over a longer timeline are needed before any definitive conclusion regarding the presence or absence of TTVs can be made. Though we find no consistent trend identifying any TTV over this ∼1 year baseline, there is better consistency amongst the higher-quality 3.5m data. TTVs can appear chaotic [5], and many transit observations are required before their behavior becomes clear, and then perhaps only after extensive effort is expended in exhausting the degenerate parameter space for possible planetary companion(s). Thus, the work presented here represents only initial efforts of the long-term study needed to investigate GL 436, and eventually other transiting M-dwarfs, in the search for TTVs.

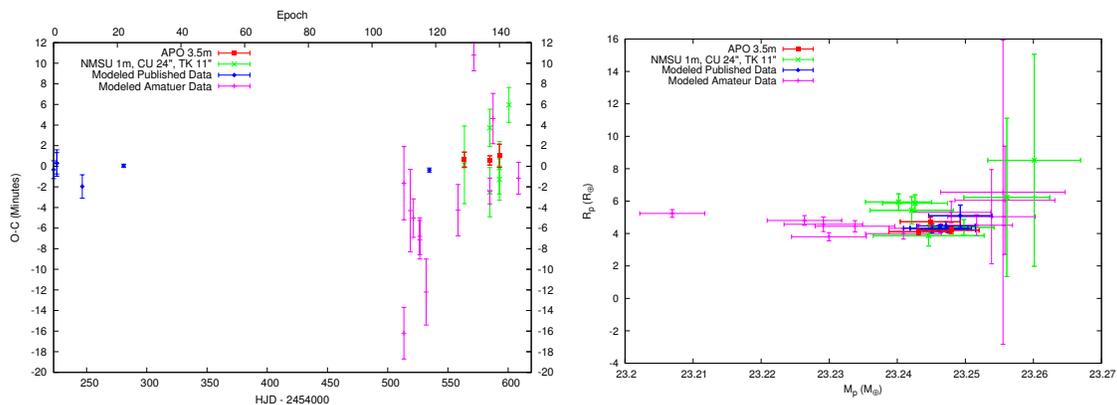

**FIGURE 3.** *Left:* Observed minus calculated (O-C) transit midpoints for all data using a newly derived ephemeris (see [4]). No variation is seen in the 3.5m APO data, though variations are seen between these and the other data; there may be some filter bandpass dependence. No clear trend in O-C is evident for TTVs <1 minute over the transit time covered (∼1 yr), and will likely require longer baselines to uncover (e.g., [4]). *Right:* Derived radius and mass of the hot-Neptune planet obtained from fitting each independent observed transit using our computed inclination for each transit. Errors displayed are $1\sigma$.

## ACKNOWLEDGMENTS

This research has been supported in part by grants from NASA, with observing time enabled through the ARC-APO 3.5m consortium, NMSU, and CU-Boulder. We are grateful to R. Alonso for providing their reduced data for analysis, the numerous amateur astronomers who provided transit observations and Bruce Gary for coordinating these.## REFERENCES

To appear in "Proceedings of the 15th Cambridge Workshop on Cool Stars, Stellar Systems and the Sun", 2009, AIP Conference Proceedings vol. 1094, ed. Eric Stempels.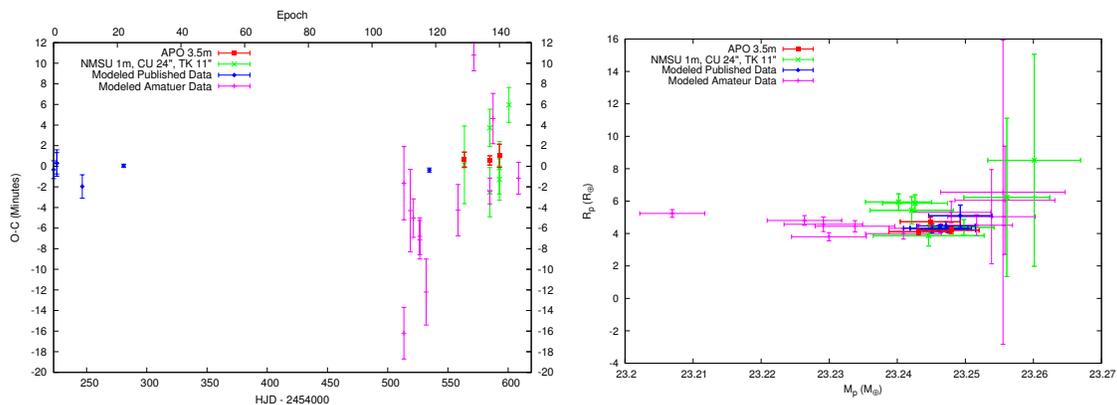

**FIGURE 3.** *Left:* Observed minus calculated (O-C) transit midpoints for all data using a newly derived ephemeris (see [4]). No variation is seen in the 3.5m APO data, though variations are seen between these and the other data; there may be some filter bandpass dependence. No clear trend in O-C is evident for TTVs <1 minute over the transit time covered (∼1 yr), and will likely require longer baselines to uncover (e.g., [4]). *Right:* Derived radius and mass of the hot-Neptune planet obtained from fitting each independent observed transit using our computed inclination for each transit. Errors displayed are $1\sigma$.

absence of TTVs can be made. Though we find no consistent trend identifying any TTV over this ∼1 year baseline, there is better consistency amongst the higher-quality 3.5m data. TTVs can appear chaotic [5], and many transit observations are required before their behavior becomes clear, and then perhaps only after extensive effort is expended in exhausting the degenerate parameter space for possible planetary companion(s). Thus, the work presented here represents only initial efforts of the long-term study needed to investigate GL 436, and eventually other transiting M-dwarfs, in the search for TTVs.

## ACKNOWLEDGMENTS

This research has been supported in part by grants from NASA, with observing time enabled through the ARC-APO 3.5m consortium, NMSU, and CU-Boulder. We are grateful to R. Alonso for providing their reduced data for analysis, the numerous amateur astronomers who provided transit observations and Bruce Gary for coordinating these.

## REFERENCES

1. Agol, E., Steffen, J., Sari, R., & Clarkson, W. 2005, MNRAS 359, 567.
2. Alonso, R., Barbieri, M., Rabus, M., et al. 2008, submitted to A&A, arXiv:0804.3030
3. Butler, R.P., Vogt, S.S., Marcy, G.W., et al. 2004, ApJ, 617, 580.
4. Coughlin, J.L., Stringfellow, G.S., Becker, A.C., et al. 2008, submitted to ApJ Letters, arXiv:0809.1664
5. Ford, E.B. & Holman, M.J. 2007, ApJ 664, L51.
6. Gillon, M., et al. 2007, A&A, 472, L13.
7. Maness, H.L., Marcy, G.W., Ford, E.B., et al. 2007, PASP, 119, 90.
8. Miralda-Escude, J. 2002, ApJ 564, 1019.
9. Southworth, J. 2008, MNRAS, 386, 1644. (and references therein)
10. Torres, G. 2007, ApJ, 671, L65.